\documentclass[pra,twocolumn,showpacs,nobibnotes,floatfix,superscriptaddress]{revtex4}

\usepackage{latexsym,amsmath,amssymb,amsfonts,mathbbol,graphicx,color}
\usepackage{dcolumn}
\usepackage{bm}

\begin{document}

\title{Superoperator Analysis of Entanglement in a Four-Qubit Cluster State}

\author{Yaakov S. Weinstein} 
\author{Jay Feldman}
\altaffiliation[Present address:]{ Johns Hopkins University, Baltimore, MD 21201, USA} 
\author{Jacob Robins} 
\altaffiliation[Present address:]{ University of Pennsylvania, Philadelphia, PA 19104, USA} 
\author{Jason Zukus} 
\altaffiliation[Present address:]{ Princeton University, Princeton, NJ 08544, USA} 
\author{Gerald Gilbert}
\affiliation{Quantum Information Science Group, {\sc Mitre},
260 Industrial Way West, Eatontown, NJ 07224, USA}


\begin{abstract}
In this paper we utilize superoperator formalism to explore the entanglement evolution of 
four-qubit cluster states in a number of decohering environments. A four-qubit cluster 
state is a resource for the performance of an arbitrary single logical 
qubit rotation via measurement based cluster state quantum computation. We are specifically 
interested in the relationship between entanglement evolution and the fidelity with which the 
arbitrary single logical qubit rotation can be implemented in the presence of decoherence
as this will have important experimental ramifications. We also note the exhibition of 
entanglement sudden death (ESD) and ask how severely its onset affects the utilization of the 
cluster state as a means of implementing an arbitrary single logical qubit rotation. 
\end{abstract}


\maketitle

\section{Introduction}

Entanglement is a uniquely quantum mechanical phenomenon in which quantum 
systems exhibit correlations above and beyond what is classically possible. 
Entangled systems are thus an important resource for many quantum information processing 
protocols including quantum computation, quantum metrology, and quantum communication 
\cite{HHH}. In the area of quantum computation, certain entangled states play a unique 
role as the basic resource for measurement-based quantum computation. The cluster state 
in particular allows for quantum computation to proceed {\it via} single qubit measurements after 
creation of the cluster state \cite{BR1}. 

An important area of research is to understand the possible degradation of 
entanglement due to decoherence. Decoherence, stemming from unwanted interactions 
between the system and environment, is a major challenge confronting experimental 
implementations of quantum computation, metrology, and communication \cite{book}. 
Decoherence may be especially detrimental to highly entangled states \cite{Dur}
and, indeed, much work has been done on studying the effects of 
decoherence on cluster states \cite{CD}. 

An extreme manifestation of the detrimental effects of decoherence on entangled states 
is ``entanglement sudden death" (ESD), in which entanglement within a system
is completely lost in a finite time \cite{DH,YE1} despite the fact that the loss of system coherence 
is asymptotic. This aspect of entanglement has been well explored in the case of 
bi-partite systems and there are a number of studies looking at ESD in multi-partite 
systems \cite{SB,ACCAD,LRLSR,YYE,QECESD,YSW} including the four qubit cluster state \cite{YSW2}. 
In addition, there have been several initial experimental ESD studies \cite{expt}. 

In this paper we study the entanglement evolution of a four-qubit cluster state 
which can be used as the basic resource to perform an arbitrary single (logical) qubit rotation 
via cluster state quantum computation. We analyze the effects of various
decoherence models on the entanglement of the pre-measurement state and compare the entanglement 
behavior to the accuracy with which the decohered state can be used to implement the desired 
arbitrary single qubit rotation. To completely characterize the  
effects of decoherence we make use of superoperator representations and 
aspects of quantum process tomography. Quantum process tomography is an experimental 
protocol which is used to completely determine (open) system dynamics. The information 
gleaned from quantum process tomography can, in turn, be used to determine a wealth of 
accuracy measures. One would expect that the proper working of cluster state based quantum 
computation would be strongly dependent on the amount of entanglement present in the pre-measurement 
cluster state. Thus, an explicit analysis of the strength of this dependence, especially
when attempting to perform basic computational gates, is essential for experimental implementations 
of cluster state quantum protocols.

A secondary aim of this paper is to analyze the effect of ESD on the implementation of 
the single logical qubit rotation. The ESD phenomenon is of interest on a fundamental level and 
important for the general study of entanglement. However, it is not yet clear what the effect of ESD is
on quantum information protocols. Are different quantum protocols helped, hurt, 
or left intact by ESD? Previous results suggest a possible connection between the 
loss of certain types of entanglement in the four qubit cluster state and the fidelity with
which measurements on the four qubit state will lead to the desired state on the remaining, 
unmeasured qubits \cite{YSW2}. 
The current paper expands these results by exploring additional decoherence mechanisms and calculating
state independent accuracy measures such as the gate fidelity. Other explicit studies of the effect of 
ESD on quantum information protocols include the three-qubit phase
flip code, a four qubit decoherence free subspace and a three qubit noiseless subsystem 
\cite{YSW,YSW3}. None of these studies find a correlation between the accuracy of 
the protocol implementation and the advent of ESD. 

\section{Cluster States}

The cluster state \cite{BR3} is a specific type of entangled state that can be used 
as a resource for measurement-based quantum computation \cite{BR1}. 
A cluster state can be constructed by rotating all qubits into the state 
$|+\rangle = \frac{1}{\sqrt{2}}(|0\rangle + |1\rangle)$ and applying 
control phase (CZ) gates, $\rm{diag}(1,1,1,-1)$, between desired pairs. In a graphical picture 
of a cluster state, qubits are represented by circles and pairs of qubits that have
been connected via CZ gates are connected by a line. A cluster state
with qubits arranged in a two-dimensional lattice, such that each (non-edge) qubit
has been connected {\it via} CZ gates with its four nearest neighbors, suffices for 
universal QC.

After constructing the cluster state, any quantum computational algorithm
can be implemented using only single-qubit measurements performed along axes in 
the $x$-$y$ plane of the qubit, {\it i.e.} the plane spanned by 
$|+\rangle = \frac{1}{\sqrt{2}}(|0\rangle+|1\rangle$),
$|+i\rangle = \frac{1}{\sqrt{2}}(|0\rangle+i|1\rangle$). 
These processing measurements are performed by column, from left 
to right, until only the last column remains unmeasured. The last column 
contains the output state of the quantum algorithm which can be extracted 
by a final readout measurement. One can view each row of the 
cluster-state lattice as the evolution of a single logical qubit in time.
Two (logical) qubit gates are performed via connections between two rows of 
the cluster state. CZ gates in particular are `built-in' to the cluster state
and simple measurement on two connected qubits in different rows automatically 
implements the gate. 

Measurement of a physical qubit in the cluster state at an angle $\phi$ from the 
$x$-axis in the $x$-$y$ plane implements a rotation on the logical qubit given by
$X(\pi m)HZ(\phi)$, where
$H =
\frac{1}{\sqrt{2}}
\left( 
\begin{array}{cc}
1 & 1 \\
1 & -1 \\
\end{array}
\right)$ 
is the Hadamard gate, and $Z(\alpha)$ ($X(\alpha$)) 
is a $z$- ($x$-) rotation by an angle $\alpha$ \cite{BR2}. The dependence of the
logical operation on the outcome of the measurement is determined by the value of 
$m = 0, 1$ for measurement outcome $-1, +1$, respectively. An arbitrary single logical qubit 
rotation can be implemented via three such measuremnts yielding:
\begin{equation}
HZ(\theta_1+\pi m_{\theta_1})X(\theta_2 + \pi m_{\theta_2})Z(\theta_3 + \pi m_{\theta_3}), \nonumber
\end{equation}
where $(\theta_1, \theta_2, \theta_3)$ are the Euler angles of the rotation. As 
an example, by drawing the Euler angles according to the Haar measure, a
random single-qubit rotation can be implemented.

We explore an arbitrary single (logical) qubit cluster-based rotation performed on an
arbitrary initial state in a decohering environment. To construct the relevant cluster, a qubit is placed 
in the desired initial state $|\psi_{in}(\alpha,\beta)\rangle = \cos\alpha|0\rangle+e^{i\beta}\sin\alpha|1\rangle$,
where $\rho_{in}(\alpha,\beta) = |\psi_{in}(\alpha,\beta)\rangle\langle\psi_{in}(\alpha,\beta)|$.
Three additional qubits (numbered 2-4) are rotated into the $|+\rangle$-state and CZ gates are then 
applied between the original qubit and 2, 2 and 3, and 3 and 4. The four qubit initial (pure) 
state is thus 
$|\psi_{4I}(\alpha,\beta)\rangle = CZ_{34}CZ_{23}CZ_{12}(|\psi_{in}(\alpha,\beta)\rangle\otimes|+\rangle^{\otimes3})$ 
or $\rho_{4I}(\alpha,\beta) = |\psi_{4I}(\alpha,\beta)\rangle\langle\psi_{4I}(\alpha,\beta)|$.

\section{Entanglement Measures}

To quantify and monitor entanglement in the above constructed types of cluster states 
as they undergo decoherence we use an entanglement measure known as the negativity, 
$N$, defined as the most negative eigenvalue of the parital transpose of the 
system density matrix \cite{neg}. There are a number of inequivalent forms of the 
negativity for any four qubit system: the partial transpose may be taken with respect 
to any single qubit, $N^{(j)}$, or the partial transpose may be taken with respect to 
any two qubits: $N^{(j,k)}$. The negativity thus defined does not differentiate different types of 
entanglement. Furthermore, due to the possible presence of bound entanglement, the disappearance
of all measureable negativity does not guarantee that the state is separable. However, the presence
of negativity does ensure the presence of distillable entanglement in the system. 

A method of monitoring specifically four qubit cluster type 
entanglement is {\it via} the expectation value of the state with respect to an appropriate 
entanglement witness \cite{EW}. Entanglement witnesses are observables with 
positive or zero expectation value for all states not in a specified 
entanglement class and a negative expectation value for at least one state of
the specified entanglement class. Entanglement witnesses may allow for an efficient,
though imperfect, means of experimentally determining whether entanglement is present 
in a state (as opposed to inefficient state tomography). This is especially important 
for experiments with any more than a few qubits as it may be the only practical means 
of deciding whether or not sufficient entanglement is present in the system.  The entanglement 
witnesses used here are specifically designed to detect four qubit cluster type entanglement 
of the kind exhibited by states of the form $\rho_{4I}(\alpha,\beta)$. 
In Ref.~\cite{TG} an entanglement witness is constructed for a  
cluster state in which the first qubit is $|+\rangle$. It is given
by $\mathcal{W}_+ = \openone/2-\rho_{4I}(\pi/4,0)$. For the current study we modify this witness
by a phase rotation of angle $\beta$ on the first qubit yielding witnesses of the form:
\begin{equation} 
\mathcal{W}_\beta = \openone/2-e^{-i\beta\sigma_z^1/2}\rho_{4I}(\pi/4,0)e^{i\beta\sigma_z^1/2},
\end{equation} 
where $\sigma_z^k$ is the Pauli $z$ spin operator on qubit $k$ and $\beta$ is the phase of 
the initial state $|\psi_{in}(\alpha,\beta)\rangle$. This witness more accurately
determines whether the cluster states of interest in this work have any four-qubit 
cluster entanglement. 

\section{Superoperator Representation}

We would like to completely describe the evolution of the single logical qubit
undergoing an arbitrary cluster-based rotation in the presence of decoherence. 
To do so we need to account for both the decoherence and the measurements of the (three) physical qubits.
For this study we assume that there is no interaction between the qubits of the 
cluster state (beyond the initial conditional phase gates used to construct 
the cluster state).  We further assume that all decoherence occurs after 
construction of the cluster state but before measurements. Measurement is 
done on each of the first three qubits in bases at angle $\theta_i, i = 1,2,3$, 
from the positive $x$-axis. As noted above, the measurement bases are chosen 
so as to implement the desired logical qubit rotation. After measurement the final state of 
the logical qubit resides on the fourth, 
unmeasured, physical qubit and is a function of the initial state of the logical qubit 
$|\psi_{in}(\alpha,\beta)\rangle$, the decoherence strength, $p$, and the three measurement angles, $\theta_i$:
$\rho_{out}=\rho_{out}(\alpha,\beta,p,\theta_1,\theta_2,\theta_3)$. 

To construct the dynamical superoperator of the one qubit logical 
gate we follow the method described in \cite{book,QPT}. We construct the 
appropriate cluster, apply decohering evolution, and perform the desired
measurements on a set of states, $|\psi_{4I}(\alpha,\beta)\rangle$, which 
span the one qubit Hilbert space (Hilbert space dimension $N = 2$). From this 
we can construct the $N^2\times N^2$ Liouvillian superoperator, $S$, where
\begin{equation}
S(p,\theta_1,\theta_2,\theta_3)\rho_{in}(\alpha,\beta) = \rho_{out}(\alpha,\beta,p,\theta_1,\theta_2,\theta_3).
\label{super}
\end{equation}
Note that in Liouvillian space, density matrices are column vectors of dimension $N^2\times 1$. 
From the superoperator $S$ we can construct the corresponding $N\times N$ Kraus operators following
\cite{Tim}. An analysis of the Kraus operator representation of the scenarios outlined 
below is done in the Appendix B. 

\subsection{Accuracy Measures}

There are two accuracy measures that we find useful for our analysis and that we use 
to compare the accuracy of the implemented gate to the evolution of the entanglement. 
These measures quantify how well the system performed the desired operation and are 
thus vital in experimental work. The first accuracy measure we utilize is the 
{\it cluster state fidelity} of the four qubit state, $\rho_{4F}(\alpha,\beta,p)$, 
before measurement but after decoherence as a function of $p$. This is given by:
\begin{equation}
F^c = Tr[\rho_{4I}(\alpha,\beta)\rho_{4F}(\alpha,\beta,p)^{\dag}].
\end{equation}
This is a simple measure which tells how close the actual final state is to the desired one 
in the presence of decoherence. The second accuracy measure is the {\it gate fidelity} of the 
attempted single (logical) qubit rotation, $U(\theta_1,\theta_2,\theta_3)$. The gate fidelity 
quantifies the accuracy with which the attempted evolution was achieved independent of the 
initial state of the system. The superoperator allows us to calculate the gate fidelity {\it via}:
\begin{equation}
F^g = Tr[S(0,\theta_1,\theta_2,\theta_3)S(p,\theta_1,\theta_2,\theta_3)^{\dag}].
\end{equation}
where 
\begin{equation}
S(0,\theta_1,\theta_2,\theta_3) = U(\theta_1,\theta_2,\theta_3)\otimes\rm{Conj}(U(\theta_1,\theta_2,\theta_3)).
\end{equation}

We next look at decohering environments of experimental interest: phase damping, 
amplitude damping and depolarization. In all three we explore the entanglement evolution as a function 
of decoherence strength assuming that the decoherence occurs prior to measurement. Measurements are 
performed on the decohered state and thus the evolution of the fidelity of the implemented operation 
can be compared to the evolution of the entanglement. Our goal is to see what correlations exist 
between entanglement degradation and the accuracy with which the cluster state can be used to implement 
the desired single logical qubit rotation. We will also note occurrences of ESD and what effect this 
phenomenon may have on the ability of the system to implement the desired rotation.

\section{Decohering Environments}

In this paper we discuss the effects of three different decohering environments on the four qubit cluster
state. They are: independent qubit phase damping, amplitude damping, and depolarization. Each of these 
environments is completely described by a Kraus operator representation. The Kraus operator representation 
for the phase damping environment is given by:
\begin{equation}
K_1 = \left(
\begin{array}{cc}
1 & 0 \\
0 & \sqrt{1-p} \\
\end{array}
\right), \;\;\;\;
K_2 = \left(
\begin{array}{cc}
0 & 0 \\
0 & \sqrt{p} \\
\end{array}
\right),
\end{equation}
for the amplitude damping environment:
\
\begin{equation}
K_1=\left(
\begin{array}{cc}
 1 & 0 \\
 0 & \sqrt{1-p}
\end{array}
\right), 
K_2=\left(
\begin{array}{cc}
 0 & \sqrt{p} \\
 0 & 0
\end{array}
\right),
\end{equation}
\
and for the depolarizing environment:
\begin{equation}
K_1=\left(
\begin{array}{cc}
 \sqrt{1-\frac{3p}{4}} & 0 \\
 0 & \sqrt{1-\frac{3p}{4}}
\end{array}
\right), 
K_{\ell} = \frac{\sqrt{p}}{2}\sigma_{\ell}, 
\end{equation}
where the $\sigma_{\ell}$ are the Pauli spin operators,
$\ell = x,y,z$. In each case we have defined a decoherence strength parameter $p$
whose exact behavior as a function 
of time is left unspecified so as to accomodate various possible experimentally relevant 
behaviors. As an example, one might have $p = 1-e^{-\kappa\tau}$ where $\tau$ is time and 
$\kappa$ is the decay constant. In the case of independent qubit dephasing, 
this decoherence behavior would decay off-diagonal terms of the density matrix
as a power of $e^{-\kappa\tau}$ and thus go to zero ({\it i.~e.}~$p\rightarrow 1$) only 
in the limit of infinite times. We also assume equal decoherence for all four qubits. 

\section{Results}

\begingroup
\squeezetable
\begin{table*}
\caption{Summary of entanglement and fidelity results for the three explored decohering environments. The 
columns show values of $p$ at which ESD is exhibited, values of $p$ where the entanglement witness cannot 
detect entanglement, gate fidelity, and the state fidelity, where $q = p-1$ and $\tilde{p} = \sqrt{1-p}$.}
\label{table}
\begin{tabular}{||c||c|c|c|c||}
\hline 
 & ESD ($N^{(j)}=0$) & ${\rm{Tr}}[\rho\mathcal{W}_{\beta}]=0$ & $F^g$ & $F^C$ \\\hline
\hline
Dephasing     & $N^{(1)},N^{(1,2)}:\;p=2(\sqrt{2}-1)$ & $p\alt .5$ & $\frac{1}{16}(10+6\tilde{p}+p(p-6\tilde{p}-7)+q(p+2\tilde{p}-2)\cos 2\theta_2$ & $\frac{1}{32}(16(1+\tilde{p})+p(p-6\tilde{p}-14)$ \\
		  & $N^{(1,3)}:\;p\simeq .938$ & & $+2q(p+2\tilde{p}-2)\cos\theta_2^2\cos 2\theta_3)$ & $-p(p-2\tilde{p}-2)\cos 4\alpha)$ \\\hline
Amplitude 	  & none  & $p < .2$ & same as dephasing & see figure \\
Damping	  & & & & \\\hline
Depolarizing  & all $N^{(j)}:\; p \le .45$ & $p \alt .2$ & $\frac{1}{8}((p-2)(-4+p(7+p(p-6)))+q^2p^2\cos 2\theta_2)$ & $\frac{1}{32}(p-2)^2(3p(3p-5)+8-qp\cos 4\alpha)$ \\\hline
\end{tabular}
\label{Tab1}
\end{table*}
\endgroup

Starting with the general state $\rho_{4I}(\alpha,\beta)$ we separately apply each of the decohering 
environments with arbitrary decoherence strength $p$ and determine the entanglement in the output 
state $\rho_{4F}(\alpha,\beta,p)$. For the depolarizing environment the eigenvalues of the partially tranposed 
states can be determined analytically. For the other environments the eigenvalues for each $\alpha$ and $p$ 
must be determined numerically. The entanglement evolution is shown in Figs. 1-3 and summarized in 
Table \ref{table}. We also calculate cluster state fidelity comparing the state before and after decoherence. 

The entanglement of the system before or after decoherence is completely independent of $\beta$. $beta$ arises 
as a rotation, $\exp[-i\frac{\beta}{2}\sigma_z]$, on the state $|\psi_{in}(\alpha,0)\rangle$. This rotation 
commutes with CZ gates and, as single qubit rotations do not effect the entanglement, the entanglement of the state 
$|\psi_{4I}(\alpha,\beta)\rangle$ is independent of $\beta$. The $\sigma_z$ rotation which gives rise to 
$\beta$ also commutes with the Kraus operators and of the phase damping and amplitude damping environments.
Thus, the entanglement of the cluster states under these types of decoherence is also independent of $\beta$. 
The depolarizing environment, however, includes (two) Kraus operators proportional to $\sigma_x$ and $\sigma_y$ 
which do not commute with the $\sigma_z$ rotation. Nevertheless, the rotation amounts to simply 
reorienting $\sigma_x$ and $\sigma_y$ and thus, taken together, the $\sigma_z$ rotation can be applied 
after the decoherence with no effect on the final state. 

\begin{figure}[t]
\includegraphics[width=4.25cm]{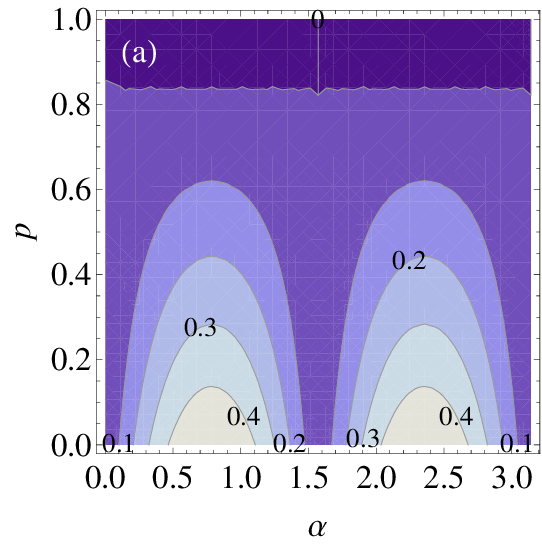}
\includegraphics[width=4.25cm]{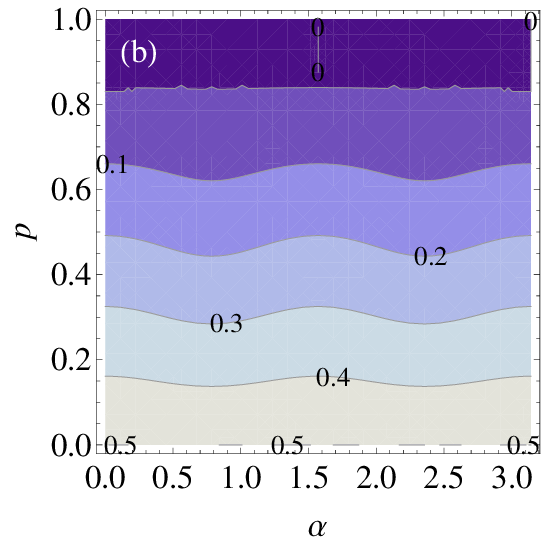}
\includegraphics[width=4.25cm]{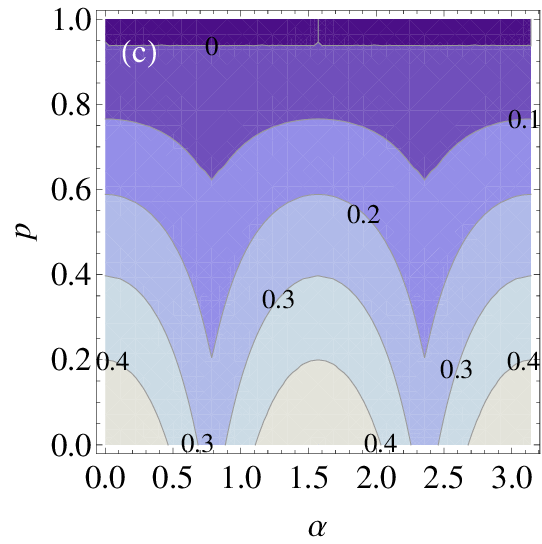}
\includegraphics[width=4.25cm]{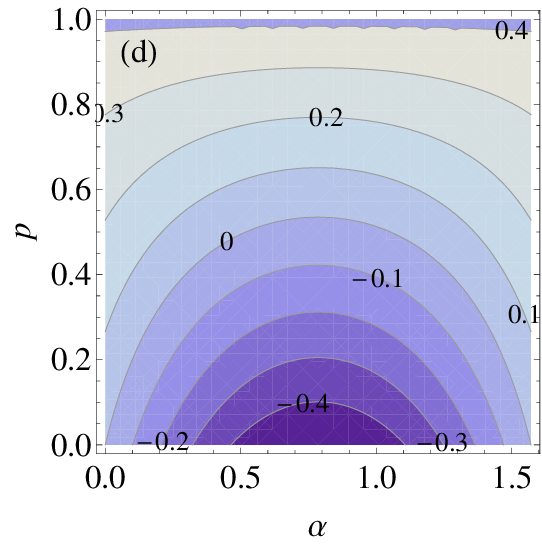}
\caption{\label{DephaseEnt} (Color online) Negativity as a function of dephasing strength, 
$p$, and initial state parameterized by $\alpha$: 
(a) $N^{(1)}$, partial transpose taken with respect to the first qubit, 
(b) $N^{(1,2)}$, partial transpose taken with respect to the first two qubits, (c) $N^{(1,3)}$ 
partial transpose taken with respect to qubits one and three. (d) Evolution of the expectation 
value of the entanglement witness as a function of initial state (the expectation value is 
not dependent on $\beta$) and decoherence strength. Notice 
that the dephasing strength at which the expectation value goes to zero is dependent
on $\alpha$ and is well below the point where ESD is exhibited for $N^{(1)}$. The four qubit 
cluster entanglement can only be observed at low levels of decoherence, $p \alt .5$. }
\end{figure}
\begin{figure}[t]
\includegraphics[width=4.25cm]{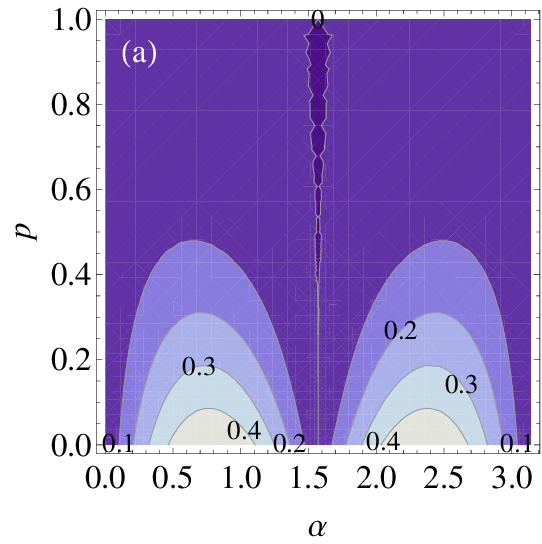}
\includegraphics[width=4.25cm]{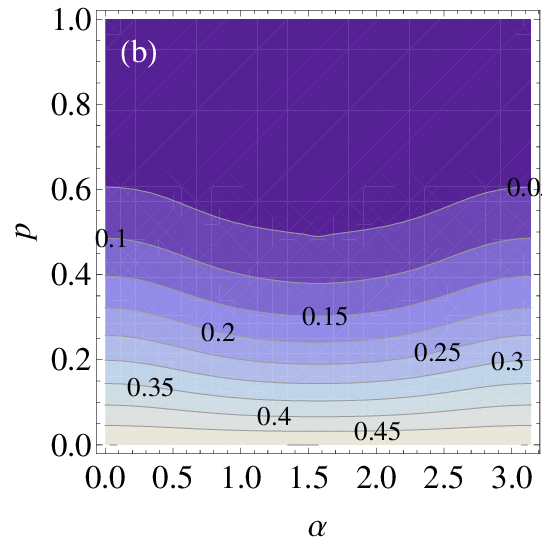}
\includegraphics[width=4.25cm]{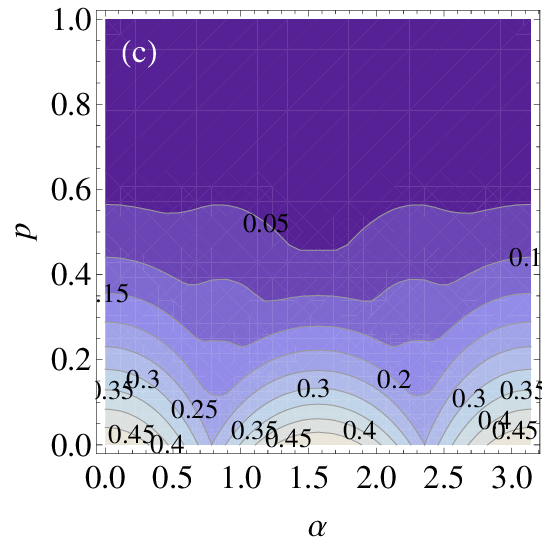}
\includegraphics[width=4.25cm]{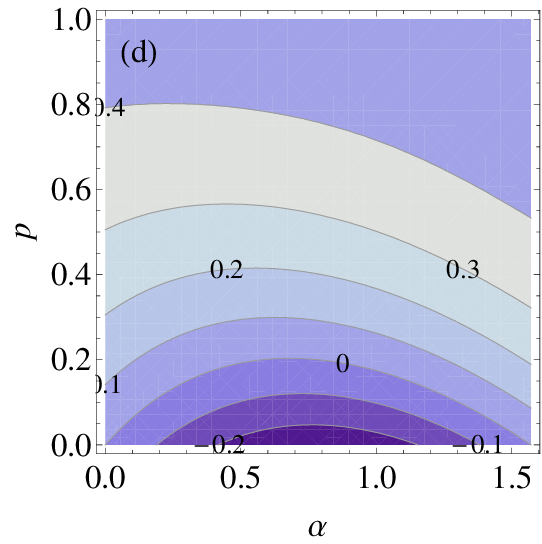}
\caption{(Color online) Evolution of various negativity measures as a function of initial state
parameterized by $\alpha$ and amplitude damping strength, $p$. 
(a) $N^{(1)}$, the entanglement goes to zero at $\alpha = \pi/2$. (b) $N^{(1,2)}$,
(c) $N^{(1,3)}$, for these measures the entanglement goes to zero only in 
the limit of $p\rightarrow 1$. (d) Expectation value of four qubit cluster state
with respect to entanglement witness $\mathcal{W}_\beta$, with $\beta = 0$, 
as a function of initial state and decoherence strength. The four qubit cluster entanglement goes
undetected at very low decoherence strengths ($p < .2$) despite the presence of some 
sort of entanglement for any non-zero $p$. }
\label{NAmp}
\end{figure}
\begin{figure}[t]
\includegraphics[width=4.25cm]{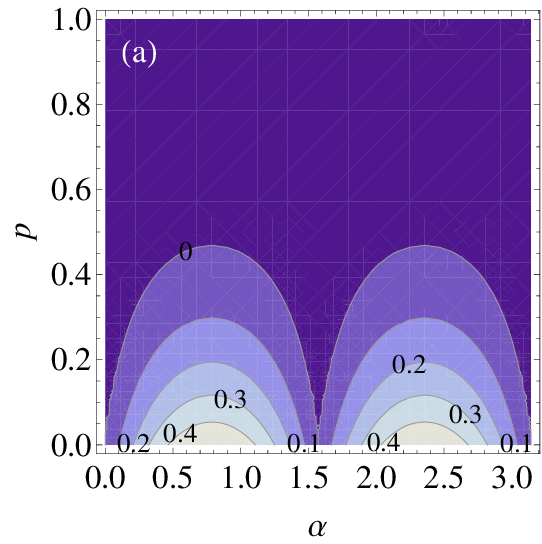}
\includegraphics[width=4.25cm]{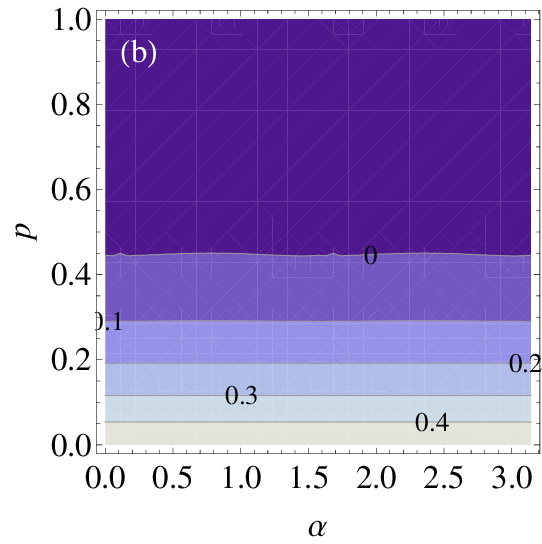}
\includegraphics[width=4.25cm]{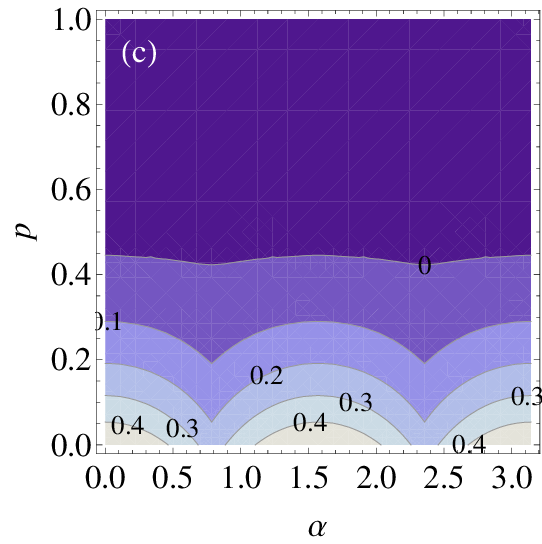}
\includegraphics[width=4.25cm]{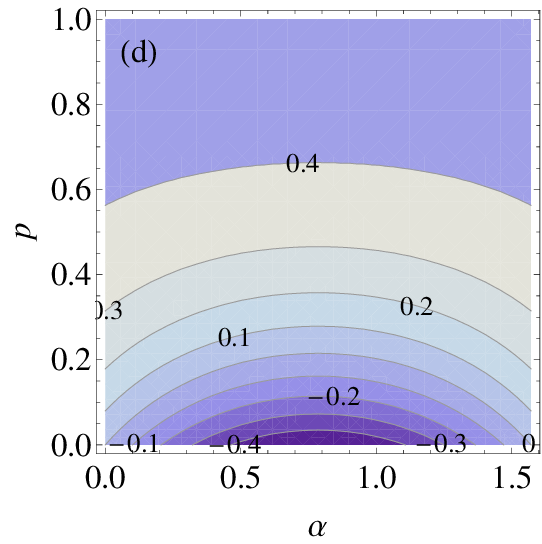}
\caption{ (Color online) Negativity as a function of depolarizing strength, $p$, and initial state 
parameterized by $\alpha$: 
(a) $N^{(1)}$,
(b) $N^{(1,2)}$, 
(c) $N^{(1,3)}$. 
(d) Evolution of expectation 
value of entanglement witness as a function of initial state and decoherence strength. Notice 
that the evolution is similar to $N^{(1)}$ though the expectation value 
goes to zero well before ESD of $N^{(1)}$ The four qubit cluster entanglement 
can only be observed at low levels of decoherence, $p \alt .2$. }
\label{DepolEnt}
\end{figure}

We then apply measurements to the first three qubits along an axis in the $x-y$ plane at angles 
$\theta_1,\theta_2,\theta_3$ from the x-zxis. From this we determine the superoperator, 
$S(p,\theta_1,\theta_2,\theta_3)$, of the attempted arbtrary rotation. These superoperators are
explicitly shown in Appendix A. 
From $S(p,\theta_1,\theta_2,\theta_3)$ we can compute the 
state independent gate fidelity as a function of the attempted rotation and the decoherence strength 
and determine the output state for any input state. Explicit equations for the two fidelity measures are 
given in Table \ref{table} and are depicted in Figs. 4-6.

There are a number of differences in the entanglement evolution between the three decohering environments.
Entanglement degrades most slowly in the dephasing environment before exhibiting ESD at high values of $p$.
In the amplitude damping environment the entanglement degrades more quickly but never exhibits ESD, and in 
the depolarizing environment the entanglement degrades most quickly and ESD is exhibited for low values of $p$.
In addition, different states lose entanglement at different rates depending on the decohering environement. For
example, $N^{(1,2)}$ in all states degrades uniformly in the depolarizing environment 
but not in the other environements. This is most likely due to uniform effect of the depolarizing 
environment over the Bloch sphere. 

With respect to detecting entanglement {\it via} the entanglement witness, 
we find for all decohering environments that the detection of four qubit cluster entanglement
goes to zero much more quickly than any of the entanglement measures (and goes to zero for the 
amplitude damping environment though no ESD is exhibited). This shows a quick demise specifically 
for the four-qubit cluster entanglement (or indicates the inefficiency of the witnesses). The maximum 
decoherence for which detection is still possible is about the same for the amplitude damping and 
depolarizing and much higher for dephasing.
 
When comparing the entanglement evolution to the evolution of the gate fidelity or 
cluster state fidelity we find only superficial correlations. In addition,  
we do not find any signature of ESD in the fidelity functions. While clearly 
both entanglement and fidelity decrease as the decoherence strength increases, 
these superficial correlations do not give rise to any problems regarding the viability
of quantum computing. We note, however, that this does not prove that ESD is completely
irrelevant with respect to quantum computation in general, it simply demonstrates that the 
effect of ESD on this specific protocol is not manifest in the fidelity measure.

The gate fidelity provides a state independent measure for the accuracy of the entire single qubit rotation 
algorithm. This fidelity, and the explicit superoperators given in Appendix A, are vital information for those 
attempting to discern the possible accuracy that can be achieved by invariably decoherent experimental systems.
We immediately note that the gate fidelity is the same in the dephasing and amplitude damping environments,
$F^g_A = F^g_z$. This is not surprising considering that the Kraus operators governing these decohering 
environments are very similar. Nevertheless, the cluster state fidelity of the two environments are 
not at all similar as seen in Figs.~4 and 5. This demonstrates the importance of utilizing multiple 
accuracy measures. In addition, the amplitude damping environment does not cause ESD for any entanglement 
measure while dephasing does. This suggests that dephasing is more harmful to the entanglement types found 
in the four qubit cluster state than is amplitude damping. Rotating the physical system
qubits, such that a dephasing environment acts as an amplitude damping environment, could conserve 
entanglement though it would not increase the accuracy of the implemented single logical qubit rotation. 
Both fidelity types decrease much more quickly in the depolarizing environment than in the other two environments. 

\begin{figure}
\includegraphics[width=4.25cm]{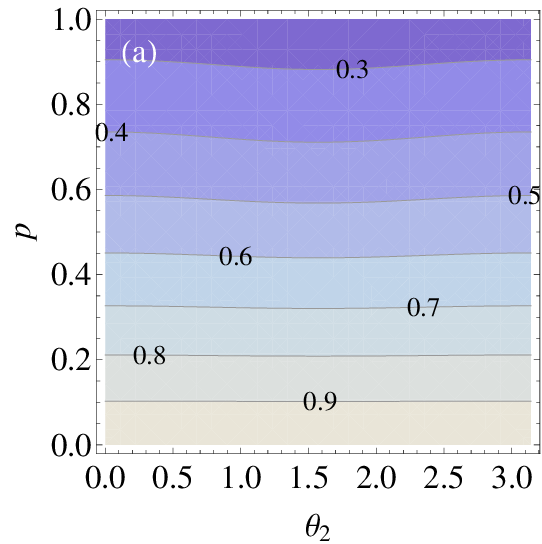}
\includegraphics[width=4.25cm]{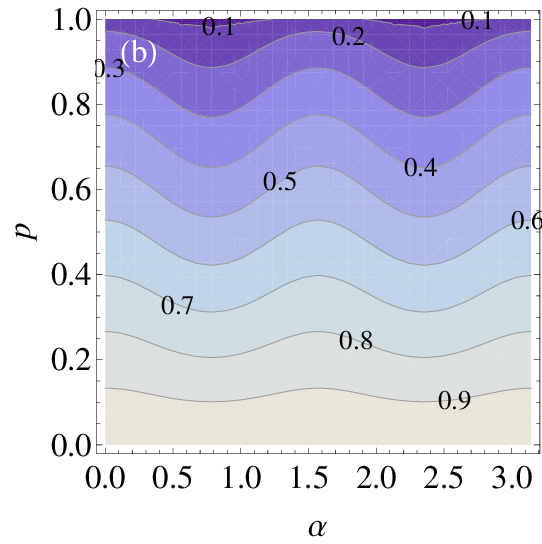}
\caption{\label{DephaseFid} (Color online) (a) Gate fidelity of an arbitrary single qubit rotation as a 
function of the rotation (parameterized by Euler angle $\theta_2$ with $\theta_3$ set to zero; 
note that the gate fidelity is independent of $\theta_1$), and dephasing strength $p$.
(b) Fidelity of pre-measurement four qubit cluster state 
as a function of initial state parameterized by $\alpha$ (this fidelity is 
independent of $\beta$) and $p$. }
\end{figure}

\begin{figure}
\includegraphics[width=4.25cm]{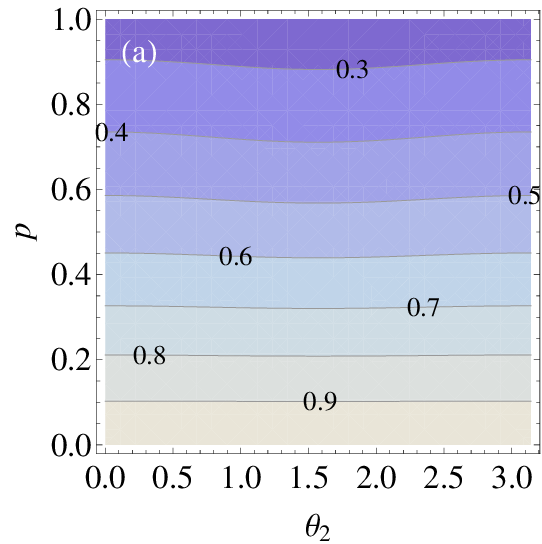}
\includegraphics[width=4.25cm]{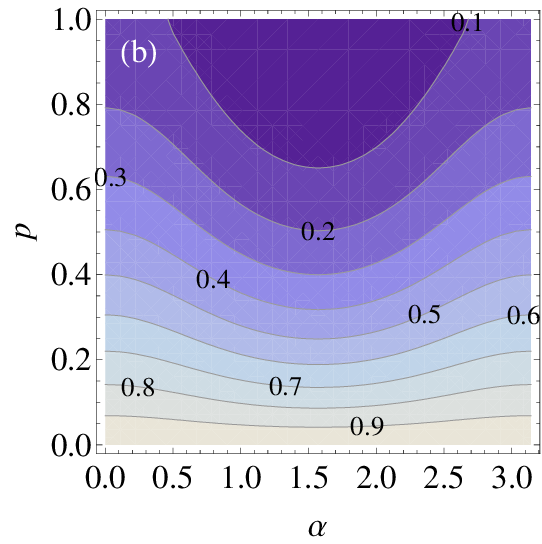}
\caption{ (Color online) 
(a) Gate fidelity as a function of amplitude damping strength, $p$ and 
choice of rotation (parameterized by $\theta_2$ with 
$\theta_3 = 0$; note that the gate fidelity is independent 
of $\theta_1$). (b) Contour plot of four-qubit state fidelity as a 
function of amplitude damping strength and initial state 
(parameterized by $\alpha$; the fidelity is independent of $\beta$). }
\label{FAmp}
\end{figure}

\begin{figure}
\includegraphics[width=4.25cm]{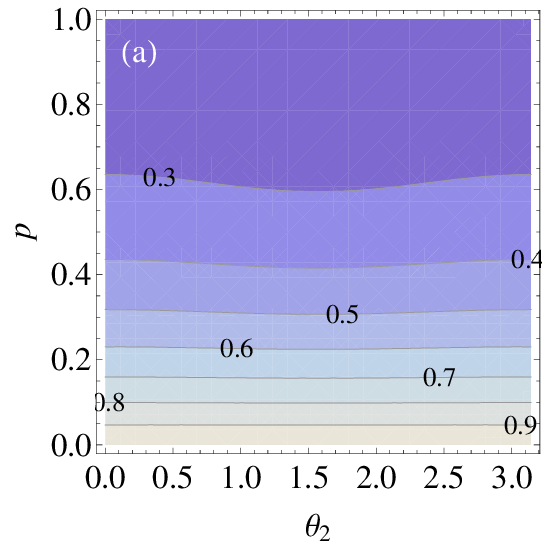}
\includegraphics[width=4.25cm]{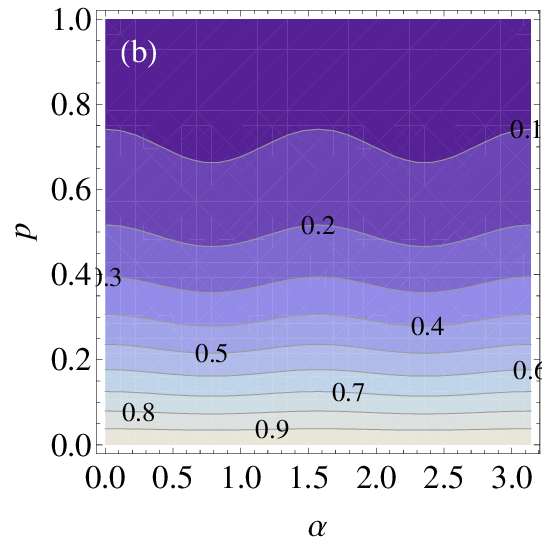}
\caption{(Color online) (a) Gate fidelity for arbitrary single qubit rotation in depolarization 
environment as a function of depolarization strength and the rotation 
(parameterized by $\theta_2$, in this case the gate fidelity does not depend 
either on $\theta_1$ or $\theta_3$). 
(b) Fidelity of four qubit cluster state as a function of initial state (parameterized
by $\alpha$ as $F^C_P$ does not depend on $\beta$) and depolarizing strength. }
\label{FDepol}
\end{figure}

\section{Conclusion}

In conclusion, we have studied an attempted implementation of an arbitrary single qubit rotation {\it via} 
cluster state quantum computation in a noisy environment. We specifically looked at three different 
decohering environments: dephasing, amplitude damping, and depolarization. Such studies are vital for 
future experiments in cluster state quantum computation as they will shed light on the types of errors 
that occur during implementation and can prescribe what types and strengths of error are tolerable
if attempting to achieve a certain accuracy of implementation. To this end, we have provided both 
accuracy measures of the implementation as a function of the attempted rotation and the decoherence type 
and strength, as well as the complete superoperator describing the entire process. The superoperators 
specifically allow for the identification of the type of error that will be manifest in the output state 
of an experiment given the error models. The accuracy measures and the superoperators are also vital for 
the determination of cluster based fault tolerance thresholds.

In addition, we have studied the entanglement evolution of the four qubit cluster state under the above 
mentioned decohering environments. This behavior is important for a number of reasons. First, multi-partite 
entanglement evolution under decoherence is still very much an area of intense study. Here we explore a 
number of negativity measures (which are bi-partite measures) and the detection capabilites
of the entanglement witness. The latter is a way to test for the presence of cluster-type entanglement in 
experiments which will be necessary before proceeding with any cluster based algorithm. 
Our results allow for comparison of entanglement evolution between the different decohering environments. 
The entanglement evolution is compared to the fidelity measures in an attempt to note any possible correspondence 
between the two. The presence or lack of a correlation between fidelity and entanglement addresses the 
general question of the role of entanglement in quantum computation. Is entanglement integral to any quantum
computation, or is it simply a byproduct of large Hilbert spaces? One may surmise that the role of entanglement
is especially vital is cluster state quantum computation as the highly entangled cluster state is the basic 
resource for any algorithm. However, we showed that there is only superficial correlation between fidelity 
and entanglement and specifically noted that the complete disappearance of entanglement upon sufficient 
decoherence, entanglement sudden death, does not have any effect of the fidelity behavior. This is especially
manifest when comparing the evolution under the dephasing and amplitude damping environments. While the gate 
fidelity of the single qubit arbitrary rotation is the same for both of these environments, the entanglement 
evolution is very different. All of this demonstrates 
that while entanglement is certainly necessary for universal quantum computation to be implemented on a cluster
state the amount of entanglement {\it per se} is not a good indicator as to how accurately the algorithm will be
implemented. 

\acknowledgments

We acknowledge 
support from the MITRE Technology Program under MIP grant \#20MSR053.

\appendix
\section{Arbitrary Single Qubit Rotation Superoperators}

In this Appendix we provide the expressions for superoperators describing the evolution of a 
single logical qubit in a cluster based quantum computation attempting to implement an arbitrary rotation 
described by Euler angles $(\theta_1,\theta_2,\theta_3)$ in the different decohering environments. 
The superoperator for the case of qubits in a dephasing environment is given by:
\begin{widetext}
\begin{equation}
S_{z} = \frac{1}{2}\left(
\begin{array}{cccc}
(1-qs2s3) & -e^{i\theta_1}q(c3-i\tilde{p}c2s3) & -e^{-i\theta_1}q(c3+i\tilde{p}c2s3) & (1+qs2s3)\\
q(c2-i\tilde{p}c3s2) & e^{i\theta_1}q(qc2c3+i\tilde{p}(s2+s3)) & -e^{-i\theta_1}q(qc2c3+i\tilde{p}(s2-s3)) & -q(c2-i\tilde{p}c3s2) \\
q(c2+i\tilde{p}c3s2) & -e^{i\theta_1}q(qc2c3-i\tilde{p}(s2-s3)) & e^{-i\theta_1}q(qc2c3-i\tilde{p}(s2+s3)) & -q(c2+i\tilde{p}c3s2) \\
(1+qs2s3) & e^{i\theta_1}q(c3-i\tilde{p}c2s3) & e^{-i\theta_1}q(c3+i\tilde{p}c2s3) & (1-qs2s3)
\end{array}
\right),
\end{equation}
where $q \equiv p-1$, $\tilde{p} \equiv \sqrt{1-p}$, and we write $\sin\theta_j$ and $\cos\theta_j$ for $j = 1,2,3$ 
as $sj$ and $cj$.  

The superoperator  for the amplitude damping environment is given by
\begin{equation}
S_{A} = \frac{1}{2}\left(
\begin{array}{cccc}
(1+p)+q^2s2s3 & e^{i\theta_1}q^2(c3-i\tilde{p}c2s3) & e^{-i\theta_1}q^2(c3+i\tilde{p}c2s3) & (1+p)-q^2s2s3 \\
q(c2-i\tilde{p}c3s2) & e^{i\theta_1}q(qc2c3+i\tilde{p}(s2+s3)) & -e^{-i\theta_1}q(qc2c3+i\tilde{p}(s2-s3)) & -q(c2-i\tilde{p}c3s2) \\ 
q(c2+i\tilde{p}c3s2) & -e^{i\theta_1}q(qc2c3-i\tilde{p}(s2-s3)) & e^{-i\theta_1}q(qc2c3-i\tilde{p}(s2+s3)) & -q(c2+i\tilde{p}c3s2) \\
-q(1+qs2s3) & -e^{i\theta_1}q^2(c3-i\tilde{p}c2s3) & -e^{-i\theta_1}q^2(c3+i\tilde{p}c2s3) & -q(1-qs2s3)
\end{array}
\right).
\end{equation}
We note that the second and third rows of the amplitude damping superoperator are exactly the same
as the second and third rows of the dephasing superoperator. 

The superoperator for the depolarizing environment is given by
\begin{equation}
S_{P} = \frac{1}{2}\left(
\begin{array}{cccc}
1-q^3s2s3 & -e^{-i\theta_1}q^3(c3+iqc2s3) & e^{-i\theta_1}q^3(-c3+iqc2s3) & 1+q^3s2s3 \\
-q^2(c2+iqc3s2) & e^{i\theta_1}q^3(qc2c3+i(s2+s3)) & -e^{-i\theta_1}q^3(qc2c3+i(s2-s3)) & q^2(c2+iqc3s2) \\
-q^2(c2-iqc3s2) & -e^{i\theta_1}q^3(qc2c3-i(s2-s3)) & e^{-i\theta_1}q^3(qc2c3-i(s2+s3)) & q^2(c2-iqc3s2) \\
1+q^3s2s3 & e^{-i\theta_1}q^3(c3+iqc2s3) & e^{-i\theta_1}q^3(c3-iqc2s3) & 1-q^3s2s3
\end{array}
\right).
\end{equation}

The above superoperators promise to be useful for experimental realizations of this cluster state protocol,
including questions of fault tolerance, as they can be used to characterize a given environment.
\end{widetext}

\section{Kraus Operator Representation}

In the main part of this paper we determined the superoperators for single logical qubit rotations 
in a cluster based quantum computer undergoing different types of decoherence. 
Recasting these superoperators in terms of Kraus operators gives additional insight 
into the evolution of the logical information under the arbitrary qubit rotation
as a function of decoherence. To calculate the Kraus operators from the superoperator 
one first determines the Choi matrix. Each Kraus operator $K_a$ is a Choi matrix 
eigenvector (unstacked so that its dimension is $N\times N$), times the square root 
of the corresponding Choi matrix eigenvalue divided by $N$ \cite{Tim}. 
We define the amplitude of a given Kraus operator, $A_a$, to be the square root of the 
Choi matrix eigenvalue divided by $N$, $A_a = \sqrt{\lambda_a/N}$.
The higher the amplitude of a Kraus operator the more significant its 
effect on the overall system dynamics. This method of Kraus operator construction 
maximizes the amplitude of one (and hence the most significant) Kraus operator. 
Using Kraus operators, the complete evolution of the system is given by
\begin{equation}
\sum_aK_a(p,\theta_1,\theta_2,\theta_3)\rho_{in}(\alpha,\beta)K_a(p,\theta_1,\theta_2,\theta_3)^{\dag} = \rho_{out}.
\end{equation}
Clearly, if there is only one Kraus operator it will be unitary with an
amplitude of 1. In this way, unitarity of the evolution 
can be quantified by $A_1$, the amplitude of the first
Kraus oprerator. In addition, the accuracy of the applied evolution can be quantified 
by the fidelity or correlation of the first Kraus operator as compared to 
the desired unitary \cite{QPT}. The fidelity is given by:
\begin{equation}
F^1 = \frac{{\rm{Tr}}[U^{\dag}K_1]}{{\rm{Tr}}[U^{\dag}U]},
\end{equation}
and the correlation is given by:
\begin{equation}
C^1 = \frac{{\rm{Tr}}[U^{\dag}K_1]}
{\sqrt{{\rm{Tr}}[U^{\dag}U]{\rm{Tr}}[K_1^{\dag}K_1]}}.
\end{equation} 
The fidelity measure accounts for both decoherent losses, a change in purity, 
and coherent errors, what we might call a change in `direction' effected by 
the first Kraus operator. The correlation is unaffected by a change in magnitude.

\begin{figure}
\includegraphics[width=4cm]{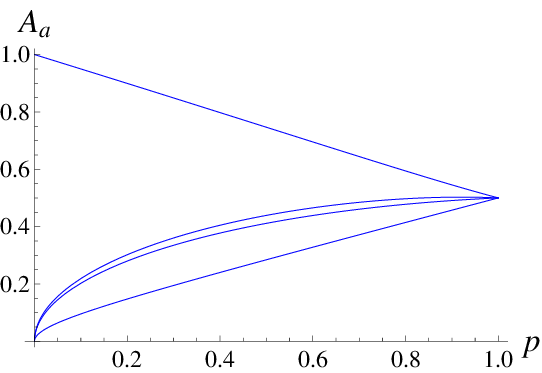}
\includegraphics[width=4cm]{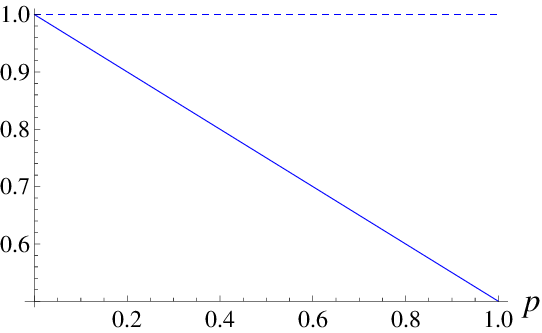}
\caption{\label{KDephase} (Color online) (left) Kraus operator amplitudes as a function of dephasing strength.
The highest Kraus operator amplitude decreases linearly until becoming
equal to the amplitudes of the other three Kraus operators. The behavior
of the highest and lowest amplitudes are independent of all measurement angles
while the middle two amplitudes depend slightly on $\theta_2$ and $\theta_3$. 
(right) Fidelity (solid line) and correlation (dashed line) of first Kraus operator 
as a function of dephasing strength.}
\end{figure}

We start with the Kraus operators of the dephasing environment.
The amplitude of the Kraus operators as a function of decoherence 
strength is shown in Fig.~\ref{KDephase}. The amplitude of the first
Kraus operator decreases linearly and the amplitude of the other three Kraus
operators increase until, at $p = 1$, the four Kraus operators have equal amplitudes. 
At that limit each of the four Kraus operator matrices have an element equal to $1/\sqrt{2}$ 
in one of the corners and all the other elements are zero. The fidelity of the first 
Kraus operator also decreases linearly with dephasing strength
while the correlation remains constant at 1 (until very high $p$) implying
purely decoherent evolution.

\begin{figure}
\includegraphics[width=4cm]{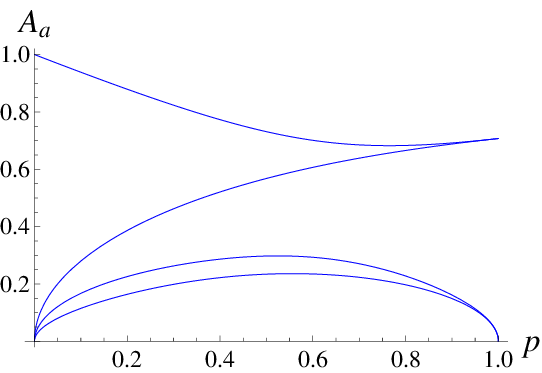}
\includegraphics[width=4cm]{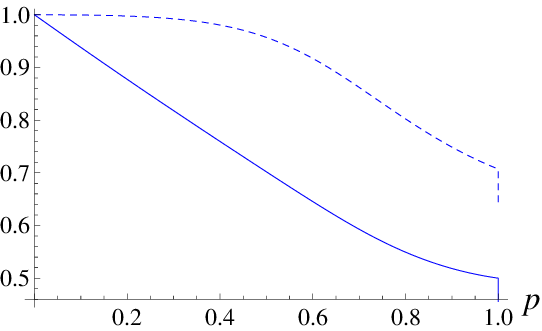}
\caption{\label{KAmp} (Color online) (left) Kraus operator amplitudes as a function of amplitude damping strength.
The behavior of all amplitudes depend slightly on $\theta_2$ and $\theta_3$. 
(right) Fidelity (solid line) and correlation (dashed line) of first Kraus operator 
as a function of amplitude damping strength. Note that in this case the correlation does
not remain constant implying a coherent affect on the dynamics of the system due to
amplitude damping.}
\end{figure}

We noted in the main part of the paper that the gate fidelities of a single logical qubit cluster state-based
arbitrary rotation in a dephasing environment and amplitude damping environment are 
the same. Nevertheless, we find that their Kraus operator representations are very different. 
Two of the Kraus operator amplitudes in an amplitude damping environment go to zero
as $p\rightarrow 1$. The remaining two Kraus operator matrices have a one 
in the upper right or left corner and zeros elsewhere. For values of $p < 1$,
the amplitude of the first Kraus operator decreases faster in the amplitude damping environment 
than in the dephasing environment. However, this descent slows as $p$ approaches one. 
The second Kraus operator always plays a more significant role in the amplitude 
damping environment than in the dephasing environment. 

The fidelity of the first Kraus operator as function of amplitude damping strength $p$ decreases linearly 
(faster than the dephasing environment) before rounding off at high values of $p$ while the correlation 
decreases, staying near one only at low values of $p$. This 
behavior, portrayed in Fig.~\ref{KAmp}, suggests that amplitude
damping, despite being decoherent dynamics, has a coherent affect on the system dynamics.
Rotating the system so that the amplitude damping acts as phase damping may increase 
the fidelity and correlation of the first Kraus operator but will not increase the 
gate fidelity of the attempted logical qubit rotation. 

\begin{figure}
\includegraphics[width=4cm]{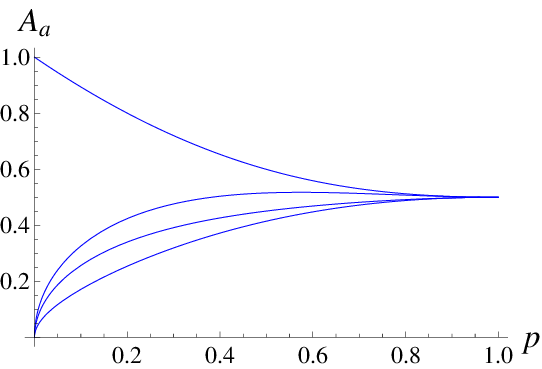}
\includegraphics[width=4cm]{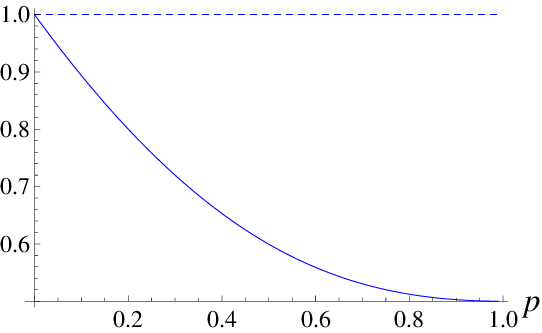}
\caption{\label{KDepol} (Color online) (left) Kraus operator amplitudes as a function of 
depolarizing strength. The behavior of all amplitudes depend slightly on $\theta_2$. 
(right) Fidelity (solid line) and correlation (dashed line) of first Kraus operator 
as a function of dephasing strength.}
\end{figure}

Of all the decohering environments studied here, the first Kraus operator amplitude decreases fastest (and not linearly) 
in a depolarizing environment. The increase of the lowest amplitude Kraus operator is also not linear. 
However, in the limit of $p\rightarrow 1$,
the depolarizing environment is like the dephasing environment in that the amplitudes of all
four Kraus operators converge to .5, as demonstrated in Fig.~\ref{KDepol}. The fidelity in a depolarzing 
environment also decreases faster than the other decoherent environments while the correlation remains 
constant at one, demonstrating that the evolution is entirely decoherent.

\end{document}